\documentclass{xray_symp}
\usepackage{graphics}
\def\ros{{\sl ROSAT }}
\def\asca{{\sl ASCA }}

\begin{document}
\title{The ULIRG NGC\,6240: Luminous extended X-ray emission and
       evidence for an AGN}
\author{Hartmut Schulz\inst{1}, Stefanie Komossa\inst{2} \and Jochen 
Greiner\inst{3}} 
\institute{Astronomisches Institut der Ruhr-Universit\"at, 44780 Bochum, 
Germany 
\and Max--Planck--Institut f\"ur extraterrestrische Physik,
 Giessenbachstra{\ss}e, 85740 Garching, Germany
\and Astrophysikalisches Institut, 14482 Potsdam, Germany}
\authorrunning{H. Schulz, St. Komossa, J. Greiner}
\titlerunning{~X-ray emission of the ULIRG NGC\,6240}
\maketitle

\begin{abstract}
We briefly review and extend our discussion of the \ros detection 
of the extraordinarily luminous ($>10^{42}$ erg/s) partly {\em extended} 
($ > 30$ kpc diameter) X-ray emission from the double-nucleus
ultraluminous infrared galaxy (ULIRG) NGC\,6240.
The \ros spectrum can be well fit by emission from
two components in roughly equal proportions: a thermal optically thin
plasma with $kT \sim$0.6 keV and a hard component that can be represented
by a canonical AGN powerlaw.
Source counts appear to have dropped by 30\% within
a year. Altogether, these findings can be well explained
by a contribution of radiation from an AGN essentially hidden at other 
wavelengths.
Fits of \asca spectra, conducted by various groups,
corroborate this result, 
revealing a high-equivalent width FeK$\alpha$ blend which 
again is straightforwardly interpreted by scattered AGN light.

If radiating at the Eddington limit, the central black hole mass does not
exceed $\sim10^7 M_{\sun}$.
We discuss implications for the formation of this AGN.
However, the luminosity in the 
remaining extended thermal component is still at the limits
of a pure starburst-wind-induced source. We suggest that the deeply
buried starburst has switched to a partially dormant phase so that
heating of the outflow is diminished and a cooling flow could have
been established. This flow may account for the extended shock heating
traced by LINER-like emission line ratios and the extremely luminous
H$_2$ 2.121 $\mu$m emission from the central region of this galaxy.
Next-generation X-ray telescopes will be able to test this proposal.
\end{abstract}

\section{Introduction}

Among known galaxies, the peculiar galaxy NGC\,6240
is outstanding in several respects: its infrared 
H$_2$ 2.121$\mu$m and [FeII] 1.644$\mu$m line luminosities
and the ratio of H$_2$ to bolometric luminosities are the
largest currently known (van der Werf et al.\ 1993). Its huge far-infrared
luminosity of $\sim 10^{12} L_{\odot}$ (Wright et al.\ 1984) comprises
nearly all of its bolometric luminosity.
Hence, owing to its low redshift of $z$=0.024,
NGC\,6240 is one of the nearest members of the class of
ultraluminous infrared galaxies 
(hereafter ULIRGs).
Its optical morphology (e.g., Fosbury \& Wall 1979, Fried \& Schulz 1983)
and its large stellar velocity dispersion of 360 km/s 
(among the highest values ever found in the center
of a galaxy: Lester \& Gaffney 1994) suggest that it is
a merging system of disk galaxies near to forming an elliptical galaxy.
Like other ULIRGs, the object contains a compact ($\sim 10^2$ pc),
luminous CO(1-0) emitting core of molecular
gas with a mass of $10^{10} M_{\odot}$ (Solomon et al.\ 1997). 
Within this core most of the ultimate
enigmatic power source of the FIR radiation appears to be hidden.
 
The most popular scenario to explain the huge IR power
is based on a superluminous
starburst (Joseph \& Wright 1985, Rieke et al.\ 1985, Heckman et al.\ 
1990).
Although starburst tracers are lacking in the optical
(Keel 1990, Schmitt et al.\ 1996)
this picture has recently been supported by ISO-SWS spectra (Lutz et al.\ 
1996).
However, our optical evidence for the presence of an 
AGN (Barbieri et al.\ 1994, 1995) supplemented by
X-ray evidence (Mitsuda 1995) suggested that a starburst
only generates part of the power. 
Subsequently, in the \ros band we found it difficult to attribute the
observed $L_{\rm 0.1 - 2.4 keV} = 10^{42-43}$ erg/s to 
starburst-induced soft X-ray generation via superwind-supershell
interaction (Schulz et al.\ 1998).
A buried AGN helps to supply part of the
X-ray luminosity and may also contribute to the FIR emission
via dust heating.  
However, even then
an appreciable remaining thermal X-ray luminosity has to be generated.
It is the purpose of the present contribution to review and extend 
our earlier discussions (Schulz et al. 1998, Komossa et al. 1998)
on these issues.
 
Observed luminosities given here were derived via
plain application of the Hubble law with
$H_0 = 50$ km s$^{-1}$ Mpc$^{-1}$ yielding a distance of 144 Mpc 
for NGC\,6240. 

\section{Observations}
The data on which our results are based were taken with the 
PSPC and HRI on board of the X-ray 
satellite {\sl ROSAT} (Tr\"umper 1983).
NGC\,6240 was observed twice with the PSPC
summing up to a total exposure time of about 8 ksec
and thrice with the HRI yielding a total time of 50 ksec.
The data were retrieved from the archive.

Data reduction procedures were carried out in a standard manner
(details are given in Komossa et al.\ 1998 and Schulz et al.\ 1998). 

\section {Results and discussion}

\subsection{Spectral fits}
\begin{figure}
\resizebox{\hsize}{!}{\includegraphics{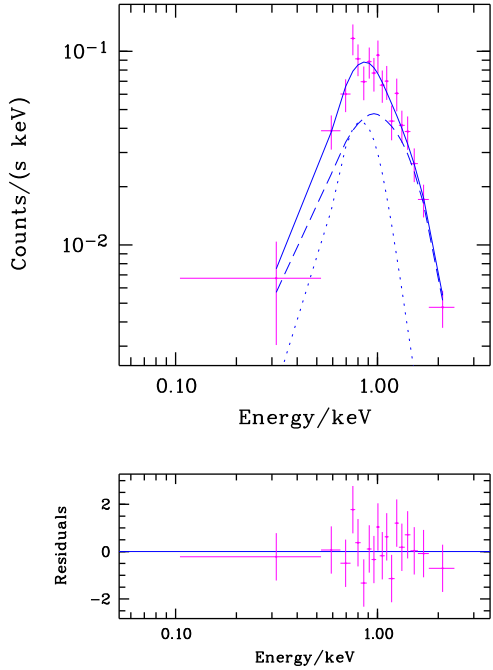}}
\vspace{-0.9cm}
\caption[]{Two-component fit to the \ros PSPC spectrum of NGC\,6240
consisting of a powerlaw (dashed line) and a Raymond-Smith model (dotted 
line).
The lower panel displays the fit residuals. }
\label{rspl}
\end{figure}

Fitting the \ros PSPC observations of NGC\,6240 we found that, 
in addition to a soft thermal component ($kT \sim 0.6$ keV), 
two-component 
fits of the (0.1--2.4) keV PSPC spectrum require a second {\em hard}
component  (Fig. \ref{rspl})
that can be represented either by {\em very hot thermal} emission 
($\simeq 
7$ keV) 
or a {\em powerlaw} with the canonical photon index $-1.9$ or slightly 
flatter.{\footnote{For
a recent discussion of the issue of 
two-component X-ray spectral models of $\sim$solar abundances in the Raymond-Smith  
component versus single-component models of very subsolar abundances 
see, e.g., Buote \& Fabian (1998) and Komossa \& Schulz (1998).}}

\begin{figure*}
\vspace{0.5cm} \hspace{1.0cm}  
 \resizebox{\hsize}{!}{\includegraphics{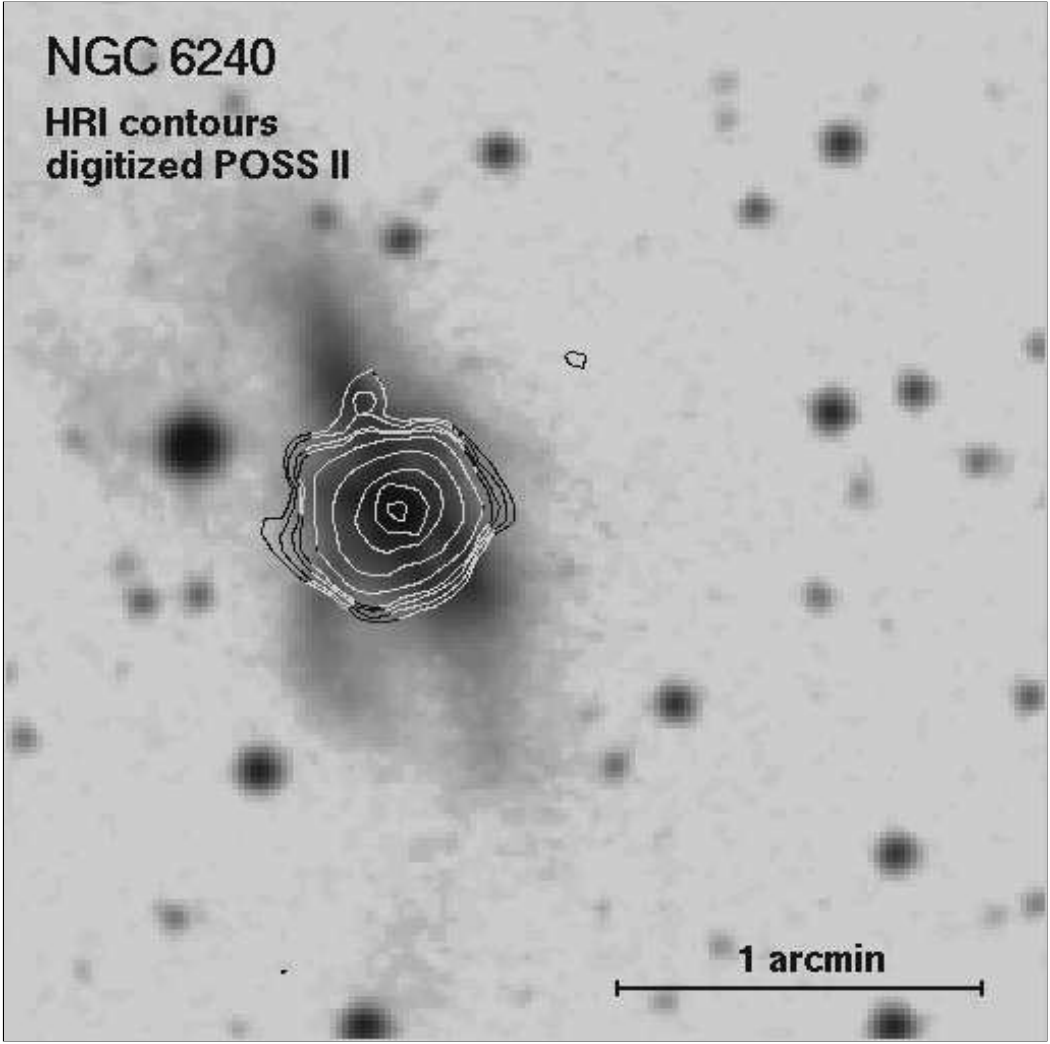}}
\vspace{-4.5cm}
\caption[]{ Overlay of the HRI X-ray contours on an optical image of 
NGC\,6240. The lowest contour is at 2$\sigma$ above the background. }
\label{ovl} 
\end{figure*}

\begin{figure}
\resizebox{\hsize}{!}{\includegraphics{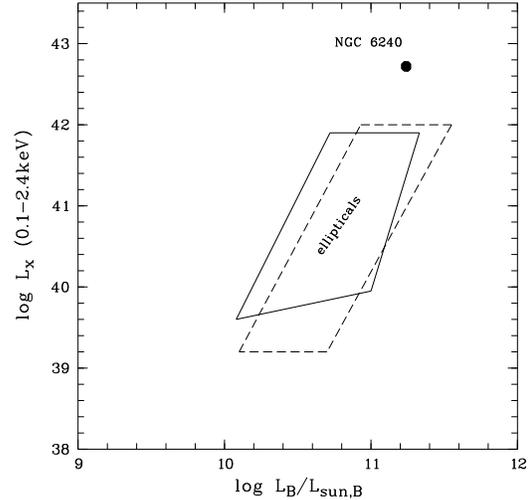}}
\vspace{-0.9cm}
\caption[]{ Locus of NGC\,6240 in the $L_{\rm x} - L_{\rm blue}$ diagram,
compared with two samples of elliptical galaxies (solid line, Canizares 
et al. 1989;
dashed, Brown \& Bregman 1998); the X-ray brightest ellipticals are those 
in the
group/cluster environment. Plotted is the total X-ray luminosity of 
NGC\,6240 (cf. Schulz et al. 1998, last row of their Tab. 2.)}
\label{lxlb}
\end{figure}

These models yield a total luminosity of {\em several} $10^{42}$ erg/s in 
the 
{\sl ROSAT} band (corroborating an earlier value given in 
Fricke \& Papaderos 1996; see also these
proceedings). To get an idea of
the lower limit for the luminosity we checked a large variety 
(even physically untenable models) of one- and
two-component spectral models (utilizing Raymond-Smith, black-body, 
bremsstrahlung, warm absorber/reflector 
and power-law models with fixed or free cold X-ray absorption). 
Taking into account the uncertainty in the distance of NGC\,6240, 
the most likely 
lower limit on the (0.1-2.4 keV) luminosity 
turns out to be $2\,10^{42}$ erg/s, with a conservative lower 
bound of $10^{42}$ erg/s. 
The HRI {\em images} reveal that part of this huge radiative power arises in a 
roughly spherical source
with strong ($\ge 2\sigma$ above background) emission out to a 
radius of 20\arcsec ($\sim 14$ kpc) 
(Fig. \ref{ovl}).
Hence, NGC\,6240 is the host of one of 
the {\em most luminous extended} X-ray sources 
in isolated galaxies (Fig. \ref{lxlb}).

The resolution of the HRI limits the contribution of a centrally 
concentrated ($< 4\arcsec$ in radius)
X-ray source to 30\% or $\sim 50$\% in flux for the spectral types 
involved here.  

\subsection{Interpretation of the hard component}
The hard component in the \ros band is of particular importance
because it has been related to scattered AGN radiation in
Komossa et al. (1998) and Schulz et al.\ (1998).
Various fits of \asca spectra (e.g., Mitsuda 1995, Kii et al.\ 1997, 
Iwasawa 
1998, Netzer et al. 1998)
revealing the extension of the hard component up to 10 keV
support this conclusion; the various fits differ in the description
of the soft component(s) and the amount of absorption of the
hard component, though. An additional feature is the 
FeK$\alpha$ blend, with a $\sim 2$ keV equivalent-width
more common for an AGN rather than a starburst. 

\begin{figure}
\resizebox{\hsize}{!}{\includegraphics{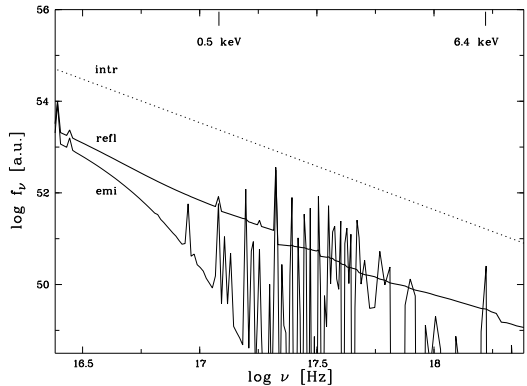}}
    \vspace{-0.3cm}
 \caption[feka]{Spectral components of a `warm scatterer', calculated
with Ferland's (1993) code {\em Cloudy}. The incident
continuum is shown as dotted line. The thin solid line corresponds to
the emitted spectrum and the thick solid line to the reflected spectrum.
The abscissa brackets the energy range 0.1 -- 10 keV.
}
\label{scat}
\end{figure}

An AGN induced hard component has to be compact which limits its
contribution to the \ros band to $\le 2.5\,10^{42}$ erg/s.
At higher energies, 
the hard component emits $L_{\rm 2-10 keV} \sim 4\,10^{42}$ erg/s
which has to be multiplied by a scattering factor $s_{\rm x}$ to get the
{\em intrinsic} X-ray luminosity $L_{\rm hx-int}$ of the hidden AGN. 
The precise efficiency and covering fraction of the mirror are highly
uncertain, but plausible models, either a {\em warm scatterer}
$-$ near-nuclear high-column-density ($N \ga 10^{23}$ cm$^{-2}$)
photoionized gas (cf. Fig. \ref{scat} and Komossa et al.\ 1998) $-$ or
$\sim 10^2$ pc extended plasma (Schulz et al.\ 1998) or a comparison
with NGC\,1068 (Ueno et al.\ 1994) suggest $s_{\rm x} \sim 10^2$,
leading to $L_{\rm hx-int} \sim 10^{44}$ erg/s, which implies for
a typical AGN continuum $L_{\rm bol}{\rm (AGN)} \sim 10^{45}$ erg/s.
Hence, the AGN contributes an appreciable fraction of the total
$L_{\rm bol}{\rm (NGC\,6240)} = 4\,10^{45}$ erg/s. With 
$L_{\rm bol}{\rm (AGN)} \sim L_{\rm Edd}$ a black hole mass of
$M_{\rm bh} \sim 10^7 M_{\sun}$ results. NGC\,6240 is expected 
to form an $L_*$ elliptical galaxy rather than a giant elliptical
after having completed its merging epoch (Shier \& Fischer 1997).
However, to match the 
relation $M_{\rm bh} \approx 0.002\, M_{\rm gal}$ (Lauer et al.\ 1997) 
for the evolved elliptical the black hole has still
to grow by an order of magnitude which requires another $10^9$ yrs of
accretion while the merger is settling down.

\subsection{Interpretation of the thermal soft component}

The \ros X-ray luminosity of the extended source of NGC\,6240 is
outstandingly large (Fig. \ref{lxlb}), even after subtracting
the hard component. Komossa et al.\ (1998) and Schulz et al.\ (1998)
found it to be at the limits for simple supernova driven 
superwind models. An additional small contribution may come from a wind
induced by the large velocity dispersion of 350 km/s (Lester \& Gaffney 
1994)
leading to shocks in the gas expelled by the red giant population.
Another interesting point is that the extended X-ray bubbles around 
elliptical
galaxies are usually brighter in the inflow phases or `when caught
in the verge of experiencing their central cooling
catastrophe' (Ciotti et al.\ 1991; Friaca \& Terlevich 1998). 
Although time scales and details for an ongoing merger are certainly
different, it is conceivable that NGC\,6240 experiences a lack of heating
when a major starburst period has ended. In this case, a cooling flow
would commence boosting $L_{\rm x}$ and presumably shock heating the ISM 
in the central kiloparsecs. Due to fragmentation, shock velocities
could be enhanced causing
the LINER like line ratios in the two nuclei (gravitational centers)
and, with lower velocities, excite the molecular cloud complex between
the nuclei
leading to the extreme H$_2$ luminosity found there (van der Werf et al.\ 
1993). 

Admittedly, so far there is not enough information to tell from a
back-of-the-envelope estimate which of the possibly competing processes
dominates (merger induced shocks or cooling flow induced shocks) but
shape and luminosity of the X-ray source suggest that more than 
a bipolar starburst outflow is going on.

\section{Conclusions}
NGC\,6240, a merger on its way to become an elliptical galaxy,
is found to harbour an exceptionally luminous extended X-ray source.
A spectral decomposition of the X-ray flux leads to a hard 
component most likely interpreted by scattered radiation from an
otherwise obscured AGN that significantly contributes to the 
FIR of the galaxy. 

As we proposed earlier, the remaining extended X-ray source 
might be explained via starburst-driven outflow, but this scenario
has been pushed to its limits and only rather detailed models could 
clarify
whether it is sufficient. A comparison with the evolution of the
extended X-ray sources around ellipticals suggests alternative 
possibilities.
E.g., we here conjectured that
the inner X-ray bubble may have currently switched to an inflow phase
which would alleviate the X-luminosity requirements and help to 
understand 
the extended central shock excited
regions seen in the optical and in the near infrared.
Whether this new picture is tenable can be further scrutinized  
with the next-generation X-ray telescopes.

\begin{acknowledgements}
St.K. and J.G. acknowledge support from the Verbundforschung under grant No. 
50\,OR\,93065 and 50 QQ 9602 3, respectively.
\end{acknowledgements}

\end{document}